\documentclass[letterpaper]{article}
\usepackage{tabls}
\usepackage{cites}
\usepackage{epsf}
\usepackage{pgffor}
\usepackage{ifthen}
\usepackage{etoolbox}
\usepackage{appendix}
\usepackage{ragged2e}
\usepackage{amsmath}
\usepackage{amssymb}
\usepackage{subcaption}
\usepackage{caption}
\usepackage[linesnumbered,ruled]{algorithm2e}
\usepackage[top=1in, bottom=1in, left=1in, right=1in]{geometry}
\usepackage{graphicx}
\usepackage{enumitem}
\setlist[itemize]{leftmargin=*}
\usepackage{hyperref}
\usepackage[nameinlink,capitalize]{cleveref}
\hypersetup{
    colorlinks=true,
    linkcolor=black,
    citecolor=black,
    urlcolor=black}

\newcommand{\EXP}[1][]{E_{#1}}

\newcommand{\BRN}[1][]{f_{\textrm{brn}_{#1}}}

\newcommand{\BOR}[1][]{\eta_{\textrm{B}_{#1}}}
\newcommand{\BORH}[1][]{\eta_{\textrm{B}_{\textrm{hist}_{#1}}}}
\newcommand{\TF}[1][]{T_{\textrm{F}_{#1}}}
\newcommand{\TFH}[1][]{T_{\textrm{F}_{\textrm{hist}_{#1}}}}
\newcommand{\TM}[1][]{T_{\textrm{M}_{#1}}}
\newcommand{\TMH}[1][]{T_{\textrm{M}_{\textrm{hist}_{#1}}}}
\newcommand{\RHOM}[1][]{\rho_{\textrm{M}_{#1}}}

\newcommand{\ROD}[1][]{\eta_{\textrm{R}_{#1}}}
\newcommand{\RODH}[1][]{\eta_{\textrm{R}_{\textrm{hist}_{#1}}}}
\newcommand{\SDC}[1][]{S_{\textrm{cool}_{#1}}}

\newcommand{\EBP}[1][]{EBP_{#1}}

\newcommand{\SIG}[1][\funcindex_g]{\Sigma_{#1}}
\newcommand{\tr}[1][]{\textrm{tr}_{#1}}
\newcommand{\fs}[1][]{\textrm{f}_{#1}}

\newcommand{\func}[1][\funcindex]{u^{#1}}
\newcommand{\funcindex}[1][\chi]{#1}
\newcommand{\var}[1][\varindex]{x^{#1}}
\newcommand{\varindex}[1][\mu]{#1}
\newcommand{\varindexone}[1][\nu]{#1}
\newcommand{\varindextwo}[1][\theta]{#1}
\newcommand{\varindexthree}[1][\epsilon]{#1}

\newcounter{Nargs}
\def\numargs#1{%
  \setcounter{Nargs}{0}%
  \foreach \x in #1{%
    \stepcounter{Nargs}%
  }%
  \the\value{Nargs}%
}  

\newcommand*{\derivative}[3][0]{%
  \def\N{\numargs{{#3}}}%
  \frac{\partial^{\ifnum1=0\N {} \else \N \fi} #2}{%
  \foreach \x in {#3} {\partial \x}}
}

\newcommand*{\uniderivative}[3][1]{%
  \frac{\partial^{\ifnum#1=1 {} \else #1 \fi} #2}{
  \partial {#3}^{\ifnum#1=1 {} \else #1 \fi}}
}

\newcommand*{\difference}[3][0]{%
  \def\M{1}%
  \def\N{\numargs{{#3}}}%
  \frac{\delta^{\ifnum1=0\N {} \else \N \fi} #2}{%
  \foreach \x in {#3} {\delta \x}}
}

\newcommand*{\unidifference}[3][1]{%
  \frac{\delta^{\ifnum#1=1 {} \else #1 \fi} #2}{
  \delta {#3}^{\ifnum#1=1 {} \else #1 \fi}}
}

\title{The Graph Theoretic Approach for \\Nodal Cross Section Parameterization}
\author{%
  \textbf{Brendan Kochunas$^1$\footnote{Corresponding Author}, Krishna Garikipati$^2$, Matthew Duschenes$^{2,3}$ and Thomas Folk$^1$} \\
  $^1$Department of Nuclear Engineering and Radiological Sciences, University of Michigan  \\
  2355 Bonisteel Blvd., Ann Arbor, MI 48109 \\ 
\\
  $^2$Department of Mechanical Engineering, University of Michigan  \\ 
    2350 Hayward Ave., Ann Arbor MI 48109 \\ 
\\
  $^3$Department of Applied Physics, University of Michigan \\
    450 Church St., Ann Arbor, MI 48109 \\
\\
  \url{bkochuna@umich.edu}, \url{krishna@umich.edu}, \url{mduschen@umich.edu},
  \url{thfolk@umich.edu}
}
\newcommand{\authorHead}{Kochunas, et al.}
\newcommand{\shortTitle}{The Graph Theoretic Approach for Nodal Cross Section Parameterization}
\begin{document}
\maketitle
\justify 

\begin{abstract}
  Presently, models for the parameterization of cross sections for nodal diffusion nuclear reactor calculations at different conditions using histories and branches are developed from reactor physics expertise and by trial and error.
  In this paper we describe the development and application of a novel graph theoretic approach (GTA) to develop the expressions for evaluating the cross sections in a nodal diffusion code.
  The GTA generalizes existing nodal cross section models into a ``non-orthogonal'' and extensible dimensional parameter space.
  Furthermore, it utilizes a rigorous calculus on graphs to formulate partial derivatives.
  The GTA cross section models can be generated in a number of ways.
  In our current work we explore a step-wise regression and a complete Taylor series expansion of the parameterized cross sections to develop expressions to evaluate them.
  To establish proof-of-principle of the GTA, we compare numerical results of GTA generated cross section evaluations with traditional models for canonical PWR case matrices and the AP1000 lattice designs.
\end{abstract}


\pagenumbering{arabic}

\section{Introduction} 
The two-step procedure for reactor analysis has been used successfully for decades, and is still a relevant methodology today \cite{Knott2010}.
A key aspect in this procedure is the generation of parameterized few-group cross sections for the downstream nodal diffusion calculations \cite{Godfrey2003}.
It is well known that current techniques for homogenization and equivalence theory provide sufficient accuracy in the nodal cross sections \cite{Smith1986}.
However, there is another, and perhaps equally important, step of  parameterizing the nodal cross sections for evaluation at the local reactor conditions.
Cross section parameterization involves the selection of a set of dependent variables that describe the space in which to parameterize the cross sections (e.g. for a PWR this would be instantaneous and historical fuel temperature, moderator temperature, and soluble boron concentration).
Next, one must determine how to construct a ``good'' case matrix i.e. what variables do you perform instantaneous perturbations (also called branches) on? What values should be chosen to perform the perturbations? How frequently should these be performed as a function of burnup?
Another outcome of the parameterization is the resulting model (or equation) for how the nodal cross sections are calculated from their parameterized form at the local reactor conditions.
Presently, the answers to questions like these, and 
the methodology for the parameterization has all been developed through well designed numerical experiments and analysis, expert judgement, and by trial and error.
Consequently, the development of a rigorous theory to systematically, and clearly, determine parameterization models may still be regarded as an open problem in reactor physics.

In this paper we describe the preliminary development and application of a novel graph theoretic approach (GTA) to derive expressions for use in nodal diffusion codes to evaluate the parameterization. 
The GTA essentially generalizes concepts and methods already used in nodal cross section models within the mathematical formalisms of graph theory.
This approach is based on the recognition that, while high-fidelity computations on complex physical systems require unknown vectors of high dimension, engineering analysis and decision-making is almost always based on quantities of interest that are typically low-dimensional functionals of the solution.
Recent work by the authors \cite{Banerjee2019} developed a novel graph theoretic approach for the representation, exploration and analysis of computed states obtained from partial differential equation (PDE) solutions of physical systems.
In this previous work, it was shown that there is a near one-to-one correspondence between the properties of PDE solutions to stationary, conservative and dissipative dynamical systems,
and the elements of graph theory.
These correspondences are via definitions, theorems and corollaries of graph theory.
Moreover, there exists a rigorous approach to develop a discrete calculus of graphs complete with differential operators such as divergence and curl, and theorems on them \cite{Hein2007,Gilboa2008,Elmoataz2008,Desquesnes2013,Lozes2014}.

The cross section models developed from a GTA can be generated in a number of ways.
In this paper we explore a step-wise regression and a complete Taylor series expansion of the parameterized cross sections to develop expressions to evaluate them.
To establish proof-of-principle of the GTA we compare numerical results of GTA generated cross section evaluations with traditional models for canonical PWR case matrices and the AP1000 lattice designs.


\section{Preliminaries of Nodal Cross Section Parameterization} 

Cross section parameterization for nodal codes involves choosing the relevant state variables, developing a case matrix, and subsequent expression for evaluating the parameterized data.
This parameterization will vary by reactor type and for a given reactor the type of analysis being performed (e.g. operational depletion or transient accident analysis) may also lead to different parameterizations.
In some cases, these parameterizations may also undergo revision for a particular plant or cycle of a given reactor type.

All parameterizations include a few common characteristics.
First, the choice of state variables for parameterization is often straightforward for a given reactor type.
The typical state variables for a PWR from \cite{Knott2010} are given in \cref{tab:statevars}, although this may be less straightforward for novel advanced reactor concepts and designs that have not been constructed or operated.
Next is the expression of the case matrix as ``histories'' and ``branches''.
Branches are instantaneous perturbations to a state variable and histories are set conditions under which depletion (e.g. time advancement) occurs.
Further, a different case matrix is typically simulated for each fuel lattice type.
There may be anywhere from 3 to 6 or 7 fuel assembly types in a reactor, and each assembly might have anywhere from 1 to 8 or more different lattice types depending on the reactor design.
A standard case matrix for PWRs is given in \cite{Knott2010}, and the resulting cross section model is given in \cref{eq:KnottXS}.
Some example expressions for evaluating the parameterized cross sections are given in \cref{eq:KnottXS,eq:GodfreyXS,eq:ParcsXS}.
Discussion of other models will be presented in future communications.

\begin{align}
\SIG[\funcindex] ~=&~ \SIG[\funcindex,\TMH](\EXP) + \delta\SIG[\funcindex,\BORH](\EXP) + \delta\SIG[\funcindex,\TF](\EXP) + \delta\SIG[\funcindex,\TM](\EXP) \label{eq:KnottXS} \\
~+&~ \delta\SIG[\funcindex,\BOR](\EXP) + \delta\SIG[\funcindex,\ROD](\EXP) + \delta\SIG[\funcindex,\TFH](\EXP) + \delta\SIG[\funcindex,\SDC](\EXP).\nonumber \\
\nonumber \\
\SIG[\funcindex] ~=&~ \SIG[\funcindex,\TMH](\EXP) + \delta\SIG[\funcindex,\BORH](\EXP)  + \delta\SIG[\funcindex,\TF](\EXP)+ \delta\SIG[\funcindex,\TM](\EXP,\BOR) \label{eq:GodfreyXS} \\
~+&~ \delta\SIG[\funcindex,\BOR](\EXP,\ROD) + \delta\SIG[\funcindex,\ROD](\EXP,\TM) + \delta\SIG[\funcindex,\TFH](\EXP) + \delta\SIG[\funcindex,\SDC](\EXP) \nonumber \\  
~+&~ \delta\SIG[\funcindex,\RODH](\EXP) + \delta\SIG[\funcindex,\EBP](\EXP).\nonumber \\ 
\nonumber \\
\SIG[\funcindex] ~=&~ \SIG[\funcindex,\textrm{base}](\EXP) + \delta\SIG[\funcindex,\ROD] + \delta\SIG[\funcindex,\TF](\EXP) + \delta\SIG[\funcindex,\TM](\EXP) \label{eq:ParcsXS}\\
~+&~ \delta\SIG[\funcindex,\BOR](\EXP) + \delta\SIG[\funcindex,\RHOM](\EXP) + \delta^2\SIG[\funcindex,\RHOM^2](\EXP) .\nonumber
\end{align}
In \cref{eq:KnottXS}, $\delta {\SIG[\funcindex,\varindex]}$ is the change in the cross section with respect to state variable $\var$ from base condition $\SIG[\funcindex]$. The forms of \cref{eq:KnottXS,eq:GodfreyXS,eq:ParcsXS} are analogous to incomplete Taylor series expansions where some terms are omitted.

\begin{table}[htb!]
    \centering
    \caption{\bf Cross Section State Space Variables}
    \label{tab:statevars}
    \begin{tabular}{|l|l|} \hline
        State Variable &  Description            \\ \hline
        $\EXP$    &  exposure                             \\ \hline
        $\BOR$  &  instantaneous boron concentration    \\ \hline
        $\TF$   &  instantaneous fuel temperature       \\ \hline
        $\TM$   &  instantaneous moderator temperature \\ \hline
        $\RHOM$  & instantaneous moderator density \\ \hline
        $\ROD$   &  instantaneous control rod fraction   \\ \hline
        $\BRN$  & instantaneous burnup                  \\ \hline
        $\BORH$ &  historical boron concentration       \\ \hline
        $\TFH$  &  historical fuel temperature          \\ \hline
        $\TMH$  &  historical moderator temperature     \\ \hline
        $\SDC$  &  shutdown cooling                     \\ \hline
    \end{tabular}
\end{table}
\begin{figure}[htb!]
    \captionsetup*[subfigure]{position=bottom}
    \centering
    \begin{subfigure}{0.45\textwidth}
        \centering
        \includegraphics[scale=0.3]{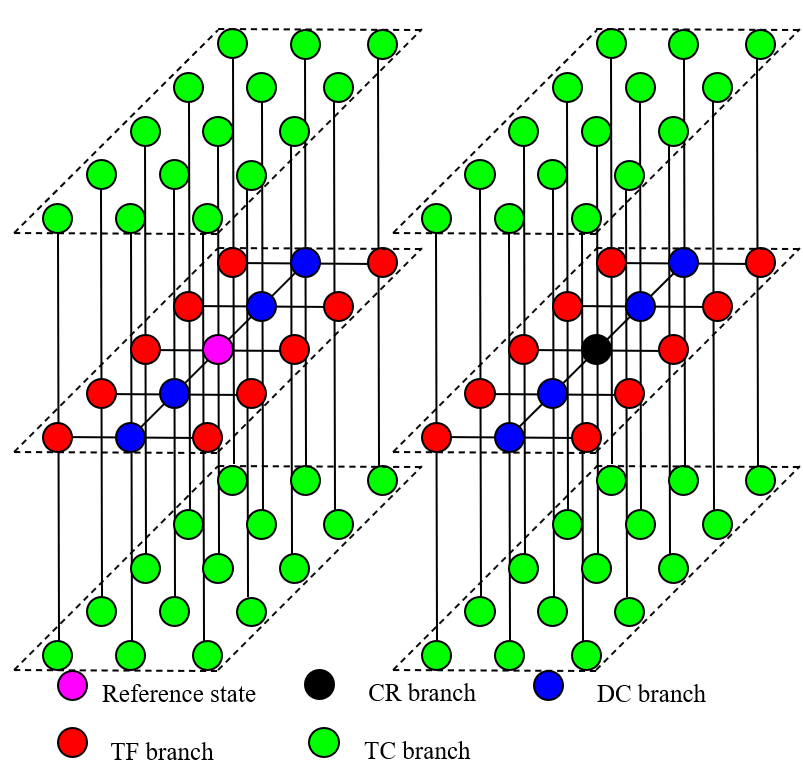}
        \caption{Conceptualization of Existing Multi-Dimensional Interpolation Representation}
    \end{subfigure}
    \begin{subfigure}{0.45\textwidth}
        \centering
        \includegraphics[scale=0.3]{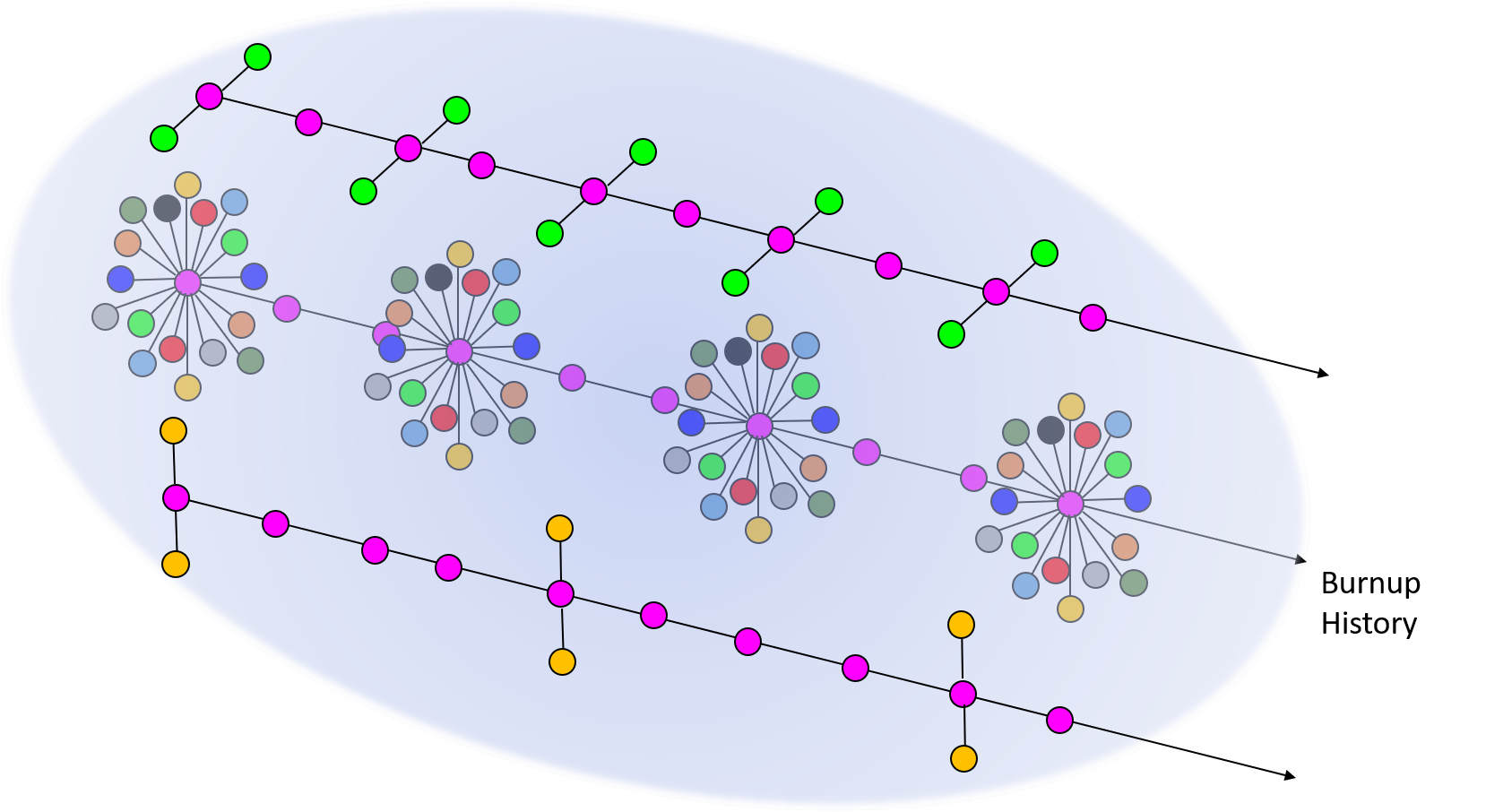}
        \caption{Conceptualization of GTA Representation}
    \end{subfigure}
    \caption{Illustrations of Parameterized Cross Section Models}
    \label{fig:xsgraphs}
\end{figure}

\section{Description of the Graph Theoretic Approach} 
The framework in which the traditional parameterized cross section models can be thought of is one in which each state variable is ``orthogonal'' to the others.
In picturing the case matrix, one often thinks of it as existing on a Cartesian grid as some n-dimensional hypercube.
This picture is very much in line with beginning to understand the case matrix as a graph.
The GTA concepts described in the next section effectively generalize this.
\Cref{fig:xsgraphs} illustrates the case matrix as this Cartesian space representation with piecewise interpolation, and the case matrix understood from a GTA where it is generalized as a discrete manifold in some low-dimensional space.


The major formalism used in this work relies on methods from graph theory. A graph, denoted by $G = (V,E)$, consists of a set of vertices, $V$, with $|V| = n$. A single vertex is denoted by  $ i \in V \in 1,\dots,n$. The vertices are connected by a set of edges, $E$. Each edge is a pair of vertices $e = (i,j) \in E$. A vertex is denoted as being the part of an edge by $i \in e$.
The graph can also have global and local attributes on the vertices and edges, as we describe below. Of importance are edge weights $w_{ij}$.
In our treatment, a state is a low-dimensional vector of functionals obtained after solving a high-fidelity computation   of dimension $K$ on a PDE, and is represented as a vertex $i \in V$. Denoting the dimension of $i$ by $k$, we have $k \ll K$.
By state transitions we mean changes in initial/boundary conditions, system parameters, or steps of the numerical solver. Thus transitions are between the vertices $i,j \in V$, and therefore they induce edges $e = (i,j)$. 

The vertices, $i \in V$ represent states $\{\var[]_i,\func[](\var[]_i)\}$, where, as introduced above, $\func[](\var[]_i)$ are quantities of interest, obtained as functionals from the high-dimensional numerical solution of a system of PDEs, and $\var[]_i$ are system parameters. The graphs naturally inherit the mathematical structure of the PDEs \cite{West2001,Newman2010}, and even confer new machinery to it \cite{Banerjee2019}.
A given state from a lattice calculation for reactor nodal cross sections  represents a vertex in the graph.
A perturbation or depletion time step off of one state to another represents an edge.

Each of the output quantities are further binned into energy groups with index $g$, and the entire set of state parameter inputs and quantity of interest outputs are:
\begin{align}
  &\textrm{Input } \var[]_i : \{\var_i\} = \{\BOR[i],\TF[i],\TM[i],\RHOM[i],\ROD[i],\BRN[i]\} \label{eq:inputs}\\
  &\textrm{Output } \func[](\var[]_i) : \{\func(\var[]_i)\} =  \{\SIG[{\tr[gi]}],~\SIG[{\fs[gi]}]\},\label{eq:outputs}.
\end{align}
Thus the graph consists of $n$ vertices $i$, each consisting of a $k$-tuple of the above quantities associated with each state: $\{\var_i,\func(\var[]_i)\}$.

Edge weights can be defined in a number of ways, and $w_{ij} = f(\var[]_j - \var[]_i)$ is natural for our purposes. It is of interest to use the graph representation of $G = (V,E)$ to develop a reduced-order model for $\func[]$ in terms of $\var[]$. Given data on a reactor in the form of the graph $G = (V,E)$, many approaches are possible. These include empirical fits, more rigorous function representations, neural networks, and dynamical system descriptions for the evolution of $\func[]$. In order to accommodate the widest range of representations, we seek to adopt a calculus on the discrete manifold that is the graph. Specifically, we adopt a discrete, non-local calculus on finite, weighted graphs \cite{Hein2007,Gilboa2008,Elmoataz2008,Desquesnes2013,Lozes2014}.

We begin by defining the weight $w = w_{ij} = 1/\vert \var[]_i - \var[]_j\vert^2$, where $\vert \var[]_i - \var[]_j\vert$ is the Euclidean norm.
Note that this gives symmetric weights and an undirected graph.
Scalars are quantities at a given state $i$, and vectors are quantities at a given state $i$, over all states $j$.
Let $\func[](\var[]_i)$ be a scalar function in this formalism for the state $i$, defined from $V \to \mathbb{R}$, and let $v(\var[]_i,\var[]_j)$ be a vector defined from $V\times V \to \mathbb{R}$. 

These functions can be thought of as mappings over the $n$ discrete data-points in $k$-dimensional space, and a calculus can be defined.
The non-local gradient operator $\nabla_w \func[]$, with respect to the weight $w$, can therefore be defined as the vector functional of scalars $\func[]$ at state $\var[]_i$, and represents the vector of derivatives with \textit{all other states} $j$:
\begin{equation}
  \nabla_w \func[](\var[]_i) \equiv (\func[](\var[]_j)-\func[](\var[]_i))\sqrt{w_{ij}},
\end{equation}
The inner product between scalars is:
\begin{align}
  \langle \func[]_1,\func[]_2\rangle \equiv&~ \sum_i \func[]_1(\var[]_i)\func[]_2(\var[]_i),
\end{align}
and the contraction, inner product, and norm between vectors are:
\begin{align}
  [v_1\cdot v_2](\var[]_i) \equiv&~ \sum_j v_1(\var[]_i,\var[]_j)~v_2(\var[]_i,\var[]_j), \\
  \langle v_1,v_2\rangle \equiv&~ \langle [v_1 \cdot v_2],1 \rangle = \sum_{ij} v_1(\var[]_i,\var[]_j)~v_2(\var[]_i,\var[]_j), \\
  |v|(\var[]_i) =&~ \sqrt{[v\cdot v](\var[]_i)}.
\end{align}
The divergence of a vector, and the Laplacian of a scalar can now be consistently defined as the scalars,
\begin{align}
  \textrm{div}_w[v](\var[]_i) =&~ \sum_j (v(\var[]_i,\var[]_j)-v(\var[]_j,\var[]_i))\sqrt{w_{ij}},\\
  \nabla^2_w[u](\var[]_i) =&~ \frac{1}{2}\textrm{div}_w[\nabla_w[u]](\var[]_i) = \sum_j (\func[](\var[]_j)-\func[](\var[]_i))\sqrt{w_{ij}}.
\end{align}
In the case of symmetric weights, there is an adjoint relation between the gradient and divergence operators:
\begin{equation}
  \langle \nabla_w u,v \rangle = - \langle u,\textrm{div}_w[v] \rangle,
\end{equation}
from which it follows that there is a divergence theorem for vectors, and the Laplacian is self-adjoint:
\begin{align}
  \sum_i \textrm{div}_w[v](\var[]_i) =&~ 0,\\
  \langle \nabla_w^2 u,u \rangle =&~ \langle u,\nabla_w^2 u\rangle.
\end{align}

Based on these non-local operators, partial derivatives with respect to $\var_i$, the $\varindex^{\textrm{th}}$ component of the $i^{\textrm{th}}$ state can be defined as the contraction between the function gradient and the specifically defined unit vector:
\begin{align}
  \unidifference{\func[](\var[]_i)}{\var_i} \approx \uniderivative{\func[](\var[]_i)}{\var_i} \equiv&~ [\nabla_w[\func[](\var[]_i)]\cdot \hat{\var[]}^{\varindex}_i] \label{eq:nonlocalpartial} \\
  =&~\sum_j (\func[](\var[]_j)-\func[](\var[]_i))(\var_j-\var_i)w_{ij}, 
\end{align}
where the unit vector at state $i$ in the $j$ direction is defined as a vector between states $i$ and $j$:
\begin{equation}
  \hat{\var[]}^{\varindex}_i \equiv \nabla_w[\var_i]
\end{equation}
Usually the weights $w_{ij}$ are chosen such that the unit vectors are normalized:
\begin{equation}
  [\hat{\var[]}^{\varindex}_i\cdot\hat{\var[]}^{\varindexone}_i] = \delta^{\varindex\varindexone},
\end{equation}
where $\delta^{\varindex\varindexone}$ is the Kronecker delta.

Higher order derivatives also can be computed using an extension of Eq. \ref{eq:nonlocalpartial}:
\begin{align}
  \difference{\func[](\var[]_i)}{\var_i,\var[\varindexone]_i} \approx \derivative{\func[](\var[]_i)}{\var_i,\var[\varindexone]_i} \equiv&~ [\nabla_w[\nabla_w[\func[](\var[]_i)]\cdot \hat{\var[]}^{\varindex}_i]\cdot \hat{\var[]}^{\varindexone}_i] \label{eq:nonlocalpartialk} \\
=&~\sum_{jk} \left((\func[](\var[]_j)-\func[](\var[]_k))(\var_j-\var_k)w_{kj} - \right. \\
&~~~~\left.(\func[](\var[]_j)-\func[](\var[]_i))(\var_j-\var_i)w_{ij}\right)(\var[\varindexone]_k-\var[\varindexone]_i)w_{ik}, \nonumber\\
\frac{\delta^k \func[](\var[]_i)}{\delta \var[\varindex_1]_i\cdots\delta \var[\varindex_k]_i} \approx \frac{\partial^k \func[](\var[]_i)}{\partial \var[\varindex_1]_i\cdots\partial \var[\varindex_k]_i} \equiv&~ [\nabla_w[\nabla_w\cdots[\nabla_w[\nabla_w[\func[](\var[]_i)]\cdot \hat{\var[]}^{\varindex_1}_i]\cdot\hat{\var[]}^{\varindex_2}_i]\cdots \cdot ]\cdot\hat{\var[]}^{\varindex_{n}}_i]\cdot\hat{\var[]}^{\varindex_k}_i] \label{eq:nonlocalpartial2}.
\end{align}
It can be shown that the above partial derivatives are, to leading order, the corresponding directional partial derivatives of differential calculus. In the limit as $n \to \infty$, and provided that $G = (V,E)$ accurately samples the topology of $\func[](x)$, convergence is attained. For finite graphs, and depending on the local topology at $\var[]_i$, however, properties such as commutativity of mixed partial derivatives do not hold.

Given these definitions for non-local derivatives, a reduced order model can be developed for systems of interest. The modelling consists of first defining the graph for the system: the distinct states of the system leading to the vertices of the graph, and which states are related through some physical process, inducing the edges of the graph. A model for these states, potentially in the form of representation functions, neural networks or differential equations can then be formulated using differential operators constructed using the above definitions.

\subsection{Step-wise Regression Methodology}

Given some basis of possible algebraic and differential operators, system inference methods can be employed to identify the reduced order model with the desired combination of expressiveness and parsimony, similar to the procedures developed by Wang et. al. \cite{Wang2019,Wang2020}.

Backwards step-wise regression consists of iteratively performing regression on an \textit{a priori} given basis of $p$ operators with parameters $\gamma$, and defining a statistical criterion to determine which operators are most, or least relevant to the model.
Operators can then be iteratively removed from the model, and further regression on this reduced basis is performed to confirm the (lack of) relevance of the operator. 
This procedure can be repeated until a minimal set of operators yields the most accurate model, and it can be posited that this minimal basis adequately represents the model.
Physical insight also can aid in determining which operators will be relevant, and help to fine tune the statistical criteria to remove or keep operators.

The statistical criteria used to reject potential operators in this communication is the \textit{F-test}, where a potential operator is removed, and the change in the regression loss is compared to the change in the model complexity.
In this communication, the root-mean-squared-error is used as the loss function between the data and the predictions, and the model complexity is defined as the number of operators.
The \textit{F-test ratio} is therefore used, with loss $L$ and complexity $C$, to determine whether to reject an operator at iteration $l$:
\begin{equation}
  \textrm{F-test}_l = \frac{\left(\frac{L^{(l)} - L^{(l-1)}}{C^{(l-1)}-C^{(l)}}\right)}{\left(\frac{L^{(l-1)}}{C^{(0)}-C^{(l-1)}}\right)}.
\end{equation}
If the ratio is less than than a tolerance $\alpha$, then the operator is said to be insignificant, and can be removed from the model.
This procedure is repeated until either further removal of operators does not yield an F-test ratio within the tolerance, or the model cannot be made more parsimonious.

The step-wise regression is therefore an algorithm with input operators and outputs data $X,y$, hyperparameters $\alpha$, and has outputs of the predicted model at each iteration in the step-wise regression $y^{(l)}$, the relevant operator parameters $\gamma^{(l)}$, as well as the statistics of the losses $L^{(l)}$ and model complexities $C^{(l)}$.
For this communication $X$ is an array of $n$ data points, corresponding to unique states in the graph, with $p$ operators of the inputs and outputs per data point.
$y$ is an array of $n$ data points, with $d$ possible dimensions of outputs. 

Several approaches are possible for reduced order models of the output quantities of interest with respect to the inputs.
These could be empirical fits, more rigorous function representations, neural networks, and dynamical system descriptions for the evolution of $\func[](x)$.
The identification of these reduced order models can be posed as system inference problems with different bases of operators for which coefficients have to be determined to satisfy expressiveness and parsimony. 

This communication will be constrained to linear regression, and it will be assumed that there is a linear relationship between the operators $X$ and outputs $y$.
Therefore $\gamma$ is an array of $p$ coefficients and the problem can be posed as finding the vector $\gamma$ such that: 
\begin{equation}
  y = X\gamma,
\end{equation}
for the matrix $X$ and vector $y$. This problem can equally be posed as minimising the mean-squared error loss:
\begin{equation}
  L = \sqrt{(y-X\gamma)^T(y-X\gamma)}.
\end{equation}
Operators are removed by fixing the associated coefficient for that operator to zero.

Due to the data being from clean direct numerical simulations, although other iterative approaches for the linear problem solutions may be used for noisy data \cite{Wang2019}, the direct pseudo-inverse (via an SVD decomposition) is used to solve each linear problem in the step-wise regression. 

There are two main schemes used for step-wise regression. Either (a) operators are considered irrelevant based on the magnitude of its coefficient, with the smallest coefficient being eliminated at each step, or (b) the operator whose removal causes the smallest increase in loss is regarded as the least significant and is eliminated.
The latter method in terms of the smallest increase in loss is used in this communication due to being less problem dependent, where some problems may have very small coefficients.

Two main approaches for step-wise regression are also considered when working with several datasets for similar systems during the reduced order modelling. In this work, the different datasets correspond to the distinct cross section types and groups First, individual regression for each given dataset, using only data and operators for that cross section set may be performed.
Second, composite regression using the concatenation of all datasets may be performed, determining the optimal generalized model of operators to keep or reject across the entire datasets throughout the step-wise procedure. These two approaches allow either individual models to be determined for each dataset, or a general model for all datasets.

Algorithm \ref{algo:step-wise_indiv} represents the scheme used when considering only one individual dataset $\funcindex$: $X = X^{\funcindex},~ y = y^{\funcindex}$, and performs the step-wise iterations only considering the rejection of operators from this individual set.
\begin{algorithm}
\KwIn{$X$, $y$, $\alpha$}
\KwOut{$\gamma^{(l)}$, $y^{(l)}$, $L^{(l)}$, $C^{(l)}$}
Fit model, given current relevant operators. \\
Reject each relevant operator individually and compute the resulting loss from the resulting models with each independent rejection. \\
Choose to reject the operator that corresponds to the minimal loss, and whose rejection also obeys the statistical criteria $\alpha$. \\
Reject chosen operator from model. \\
Calculate loss and complexity statistics, given the chosen operator is rejected. \\
Repeat procedure with the increasingly parsimonious model, until no operators can be rejected.
\caption{Individual step-wise Regression}
\label{algo:step-wise_indiv}
\end{algorithm}

Algorithm \ref{algo:step-wise_multi} represents the scheme used when considering the composite set of datasets $X = \{X^{\funcindex}\},~ y = \{y^{\funcindex}\}$, and performs the step-wise iterations considering the rejection of all operators from these sets, and chooses the optimal operator to reject. It is assumed that all individual models contain the same base set of operators.
\begin{algorithm}
\KwIn{$X$, $y$, $\alpha$}
\KwOut{$\gamma^{(l)}$, $y^{(l)}$, $L^{(l)}$, $C^{(l)}$}
Fit individual models for each group independently, given current relevant operators. \\
Reject each relevant operator independently in all groups and compute resulting losses from the resulting models with each separate rejection, as per the individual step-wise regression. \\
Find the set of operators that correspond to the minimum loss in each individual group, and obey the statistical criteria $\alpha$. \\
From this set of operators, chose the operator which has the minimum total additive loss across all groups, when that operator is removed from each group's model.\\
Reject the chosen operator from all groups. \\
Calculate loss and complexity statistics for each individual group, given the chosen operator is rejected. \\
Repeat procedure with increasingly parsimonious models, until no operators across all groups can be rejected.
\caption{Composite step-wise Regression}
\label{algo:step-wise_multi}
\end{algorithm}

The composite method chooses the order to reject operators that optimize the loss across all groups at each iteration. However the individual method may reject operators in such an order that the predicted loss at a given iteration is lower than the composite method loss at that same iteration.

\subsection{Taylor Series Approach}
In this work, we choose to specify the response function as a truncated Taylor series expanded around the ``base'' case $i=0$: $(\var[]_{0},\func[]_{0})$, with the $k^{\textrm{th}}$ order partial derivatives being $\frac{\partial^k \func[]}{\partial \var[\varindex_1]\cdots \partial \var[\varindex_k]}$, and perturbations in the $\var$ direction being $\Delta \var = \var - \var_0$:
\begin{equation}
    \func(x) = \func(\var[]_0) + \sum\limits_{\{\varindex\}} \uniderivative{\func}{\var}\bigg\vert_{\var[]_0}\Delta \var + \frac{1}{2}\sum\limits_{\{\varindex,\varindextwo\}}\derivative{\func}{\var,\var[\varindextwo]}\bigg\vert_{\var[]_0} \Delta \var \Delta \var[\varindextwo] ~+~ O(\Delta \var \Delta\var[\varindextwo] \Delta \var[\varindexthree]). \label{eq:taylor_exact}
\end{equation}
If the partial derivatives at $\var[]_0$ are exactly evaluated, the above representation would be accurate up to higher-order terms.
For finite graphs, however, the derivatives from the non-local calculus in Eq. \ref{eq:nonlocalpartialk} can only be evaluated approximately.
Even still, an approximation can be obtained to the truncated Taylor series by introducing the coefficients $\gamma^{\Xi}$, where $\Xi = \{\varindex_1,\varindex_2, \dots \varindex_k\}$, as coefficients of the $k^{\textrm{th}}$ order derivatives so that the truncated Taylor series can be approximated as 
\begin{align}
    \func(x) \approx \gamma^{0}\func(\var[]_0) ~&+~ \sum\limits_{\{\varindex\}}\gamma^{\{\varindex\}}\unidifference{\func}{\var}\bigg\vert_{\var[]_0} \Delta \var \label{eq:taylor_approx} \\ &+~\frac{1}{2}\sum\limits_{\{\varindex,\varindextwo\}}\gamma^{\{\varindex,\varindextwo \}}\difference{\func}{\var,\var[\varindextwo]}\bigg\vert_{\var[]_0} \Delta \var \Delta \var[\varindextwo] ~+~ O(\Delta \var \Delta\var[\varindextwo] \Delta \var[\varindexthree]). \nonumber
\end{align}
This Taylor series approximation is a reduced-order model in the form of a functional representation for $\func(x)$, with linear coefficients $\gamma^{\Xi}$ to be determined from case matrix data for $\var[]_i$ and $\func(\var[]_i)$ defined in Equations \ref{eq:inputs} to \ref{eq:outputs}.
These coefficients, in the case of infinite knowledge of the system, should be identically $\gamma^{\Xi} = 1$ if the Taylor series holds exactly. A verification of how well the approximated Taylor series holds can be conducted by observing which coefficients are close to $1$.

\subsection{Demonstration using AP1000 Lattices}

Westinghouse's AP1000\textsuperscript{\textregistered} PWR features an advanced core that includes five fuel regions, intra-assembly enrichment zoning, and a combination of burnable absorbers. Lattices of the active fuel region provide distinguishable characteristics that are relevant to demonstrate GTA performance. 

\begin{figure}[htb!]
    \centering
    \includegraphics[scale=1.0]{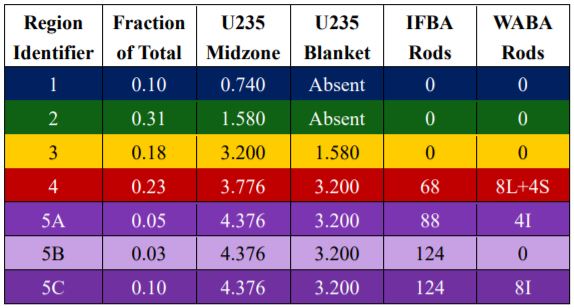}
    \caption{Fuel Summary Characteristics by Region \cite{Franceschini2014}}
    \label{fig:fuelsummary}
\end{figure}

The $^{235}$U enrichments of the fuel regions of the core span a range from natural U (0.71 wt\%) to 4.8 wt\% $^{235}$U. Of the five fuel regions, regions 1 to 3 feature a radially uniform enrichment, and regions 4 and 5 feature radial enrichment zoning. Burnable absorbers are present in regions 4 and 5 in the form of IFBA and WABA rods. A summary of the fuel region characteristics are shown in \cref{fig:fuelsummary}.

\begin{figure}[htb!]
    \centering
    \includegraphics[scale=1.0]{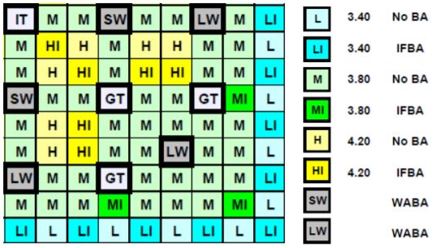}
    \caption{Region 4 Loading Pattern at the Core Axial Mid-plane \cite{Franceschini2014}}
    \label{fig:lattice_r4}
\end{figure}

\begin{figure}[htb!]
    \centering
    \includegraphics[scale=1.0]{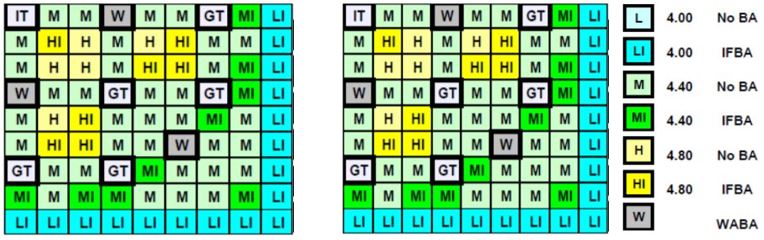}
    \subfloat[\label{lp_5A} Region 5A]{\hspace{.3\linewidth}}
    \subfloat[\label{lp_5C} Region 5C]{\hspace{.5\linewidth}}
    \caption{Region 5 Loading Patterns at the Core Axial Mid-plane \cite{Franceschini2014}}
    \label{fig:lattice_r5}
\end{figure}

The loading pattern for regions 4 and 5 lattices are illustrated in \cref{fig:lattice_r4,fig:lattice_r5}, respectively. In each image, the lattice is shown with quarter symmetry. The lattices feature three enrichment zones; low enrichment at the periphery, medium enrichment in the interior, and high enrichment towards the center. Fuel rods that contain an integral fuel burnable absorber (IFBA) are denoted with "I" following the enrichment level (L, M, H). The fuel pellets in the Westinghouse IFBA rods are coated with a ZrB$_2$ material. Annular inserts occupy the lattice slots in the form of instrument tubes (IT), guide tubes (GT), and short/long WABA tubes (SW, LW). The Wet Annular Burnable Absorber (WABA) is a type of burnable absorber that contains an $Al_2O_3-B_4C$ mixture inside of an annular tube.

Region 5 fuel consists of three lattices; two loading patterns are shown in \cref{fig:lattice_r5}. Region 5B loading pattern is the same as 5C except for the absence of WABA inserts.

The purpose of this demonstration is to establish a proof of principle for the GTA. We use a well defined case matrix \cite{Knott2010} and apply the GTA to derive new forms for the parameterized cross sections using direct numerical simulation data.
We aim to investigate whether the resulting models produced by the GTA are similar in form to established burnup dependent first-order perturbative models \cite{Knott2010}
\begin{equation}
  \SIG[\funcindex_{g}] = \SIG[\funcindex_{g}](\BRN) + \sum\limits_{\{\varindex\}} \delta {\SIG[\funcindex_{g},\varindex]} (\BRN),\label{eq:model_true}
\end{equation}
where $\delta {\SIG[\funcindex_{g},\varindex]}$ denotes the perturbation with respect to the state variable $\var$, and it is assumed that these models are cross section type and group independent. Here, all sums over the state parameters exclude the burnup state variable. 
The form of this model suggests we should model cross sections as Taylor series expansions about the case matrix base case, and about the burnup values, as per the depiction in \cref{fig:xsgraphs}, using discrete non-local calculus definitions for the non-commuting derivatives:
\begin{align}
  \SIG[\funcindex_{g}] = \gamma^{\BRN}\SIG[\funcindex_{g}](\BRN) ~&+~ \sum\limits_{\{\varindex\}} \gamma^{\{\varindex\}} \unidifference{\SIG[\funcindex_{g}](\BRN)}{\var} \Delta \var \label{eq:model_taylor} \\ &+~
  \frac{1}{2}\sum\limits_{\{\varindex,\varindextwo\}} \gamma^{\{\varindex,\varindextwo\}} \difference{\SIG[\funcindex_{g}](\BRN)}{\var,\var[\varindextwo]} \Delta \var \Delta\var[\varindextwo] ~+~ O(\Delta \var \Delta\var[\varindextwo] \Delta \var[\varindexthree]), \nonumber
\end{align}
where $\Delta \var$ represents the change in state variable from the base state, and $\unidifference{h}{x}$ represents the non-exact derivative of a function $h$ with respect to $x$ from the non-local calculus. Here, additional multiplicative coefficients $\gamma$ are included in the Taylor series in order to account for the derivatives not being exact, and must be fit for using regression methods. The base term coefficient for the burnup expansion $\gamma^{\BRN} \approx 1$ can be thought of as an estimate for the linear response due to burnup.

In this Taylor series approach, where the initial basis of operators consists of a constant base term, and up to second order derivatives, it is expected that the constant term, and then lowest order derivatives should be found to be most relevant by the backwards step-wise regression.

\section{Results} 
\begin{figure}[ht]
  \centering
  \begin{subfigure}[t]{0.49\textwidth}
  \centering
  \includegraphics[width=0.8\textwidth]{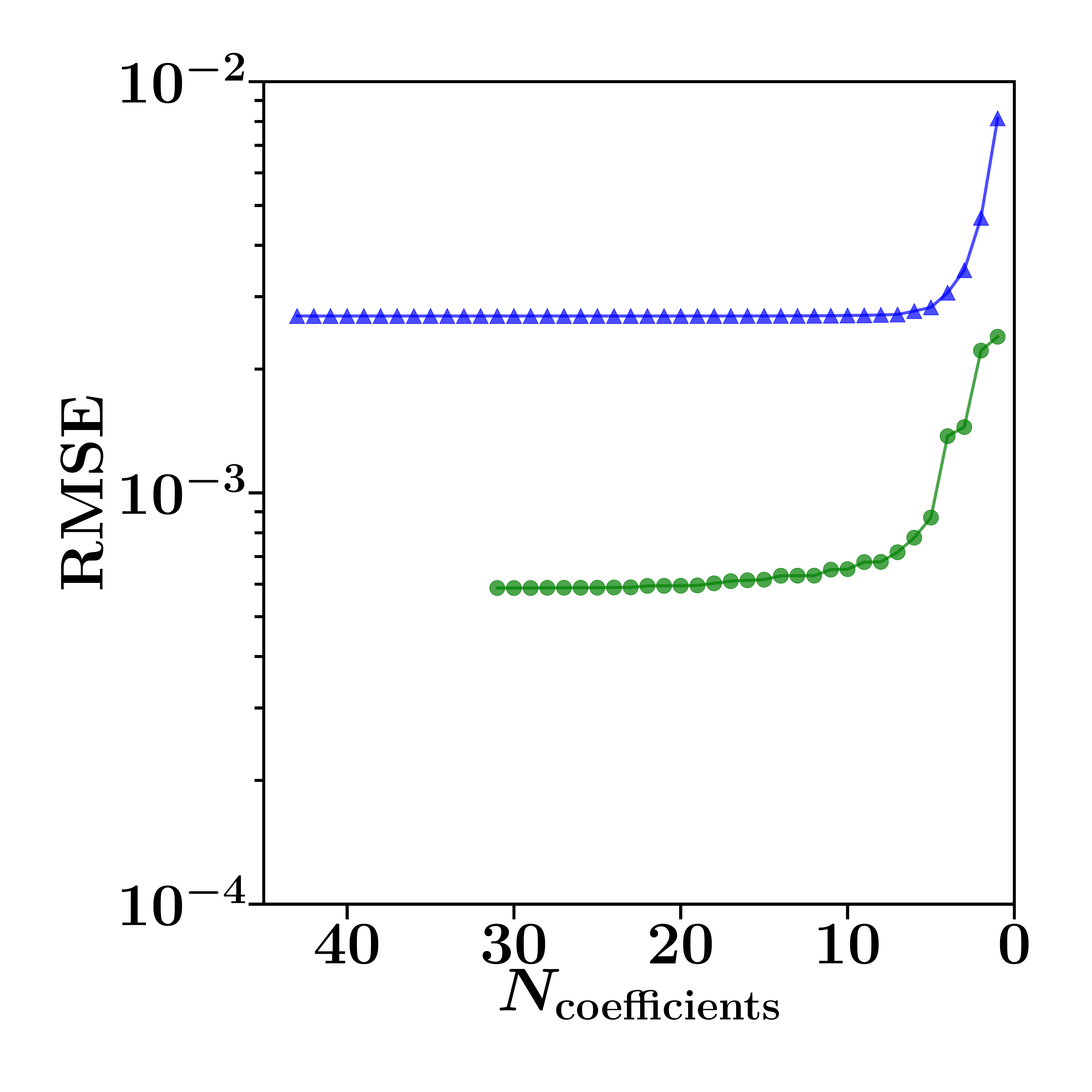}
  \subcaption{\bf Individual approach.} 
  \label{fig:Loss_r1_updated_1_XSF1_taylor_indiv_BURNUP_None}
  \end{subfigure}
  \hspace{-2cm}
  \begin{subfigure}[t]{0.49\textwidth}
  \centering
  \includegraphics[width=0.8\textwidth]{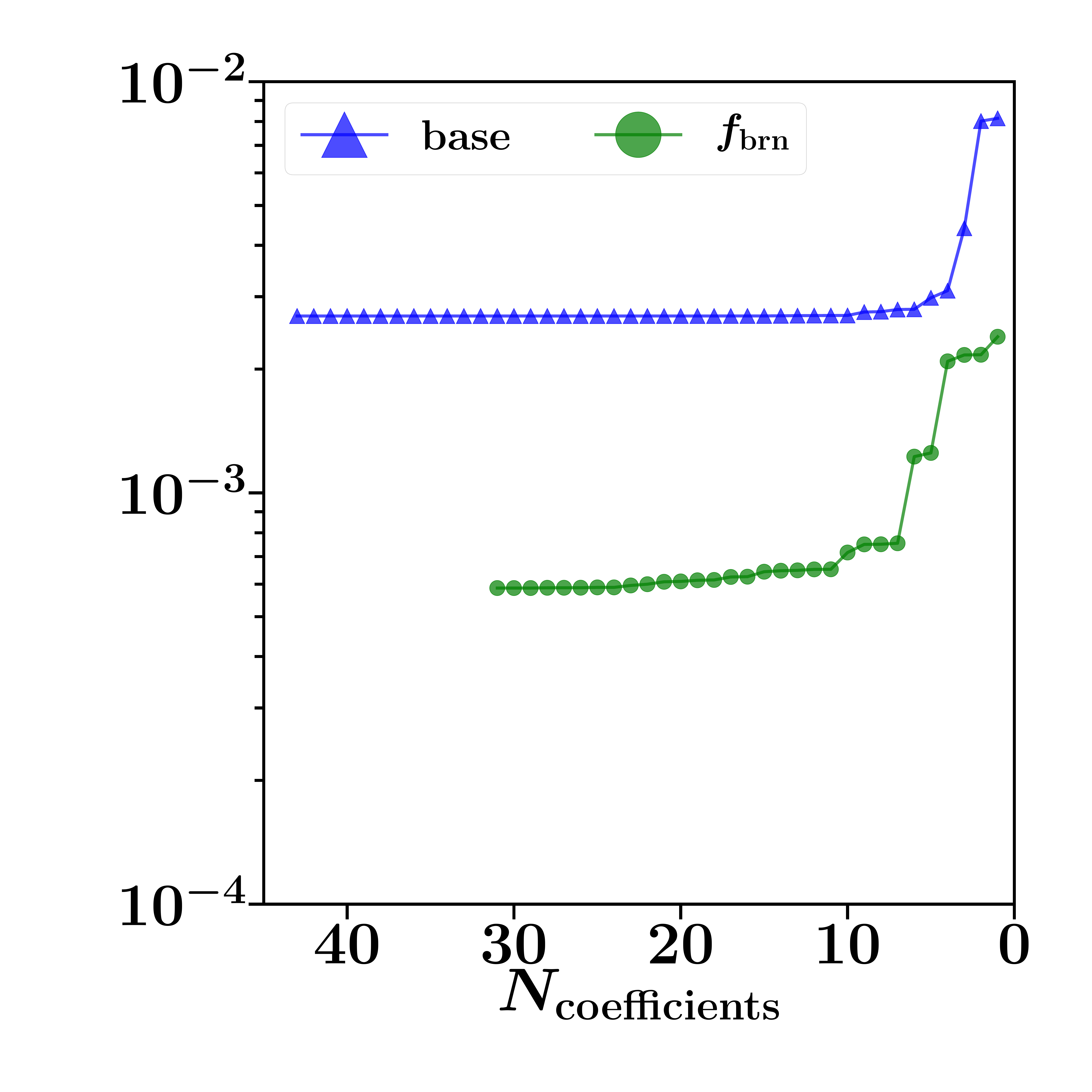}
  \subcaption{\bf Composite approach.}
  \label{fig:Loss_r1_updated_1_XSF1_taylor_multi_BURNUP_None}
  \end{subfigure}  
  \caption{\bf Root mean square loss curves between the data and cross section models over the decreasing number of operators for $\SIG[{\fs[1]}]$. The individual and composite taylor series approach, expanded around the base case and burnup states are compared.}
  \label{fig:Loss_r1_updated_1_XSF1_taylor_indiv_multi_BURNUP_None}
\end{figure}

An example set of results of the predicted model over the step-wise regression iterations, for the fission cross section, $\SIG[{\fs[1]}]$ are shown as the loss curves in \cref{fig:Loss_r1_updated_1_XSF1_taylor_indiv_multi_BURNUP_None}. Here, we consider the Taylor series expansions around the case matrix base state, as well as each burnup state along the burnup history. The latter is more indicative of current practice. We also consider the effect of performing step-wise regression either considering solely individual cross section data to form a cross section dependent model, or considering the composite data of all cross sections types and groups when performing the regression and deciding which operators form a cross section independent model. 

\begin{figure}[ht]
  \centering
  \begin{subfigure}[t]{0.49\textwidth}
  \centering
  \includegraphics[width=\textwidth]{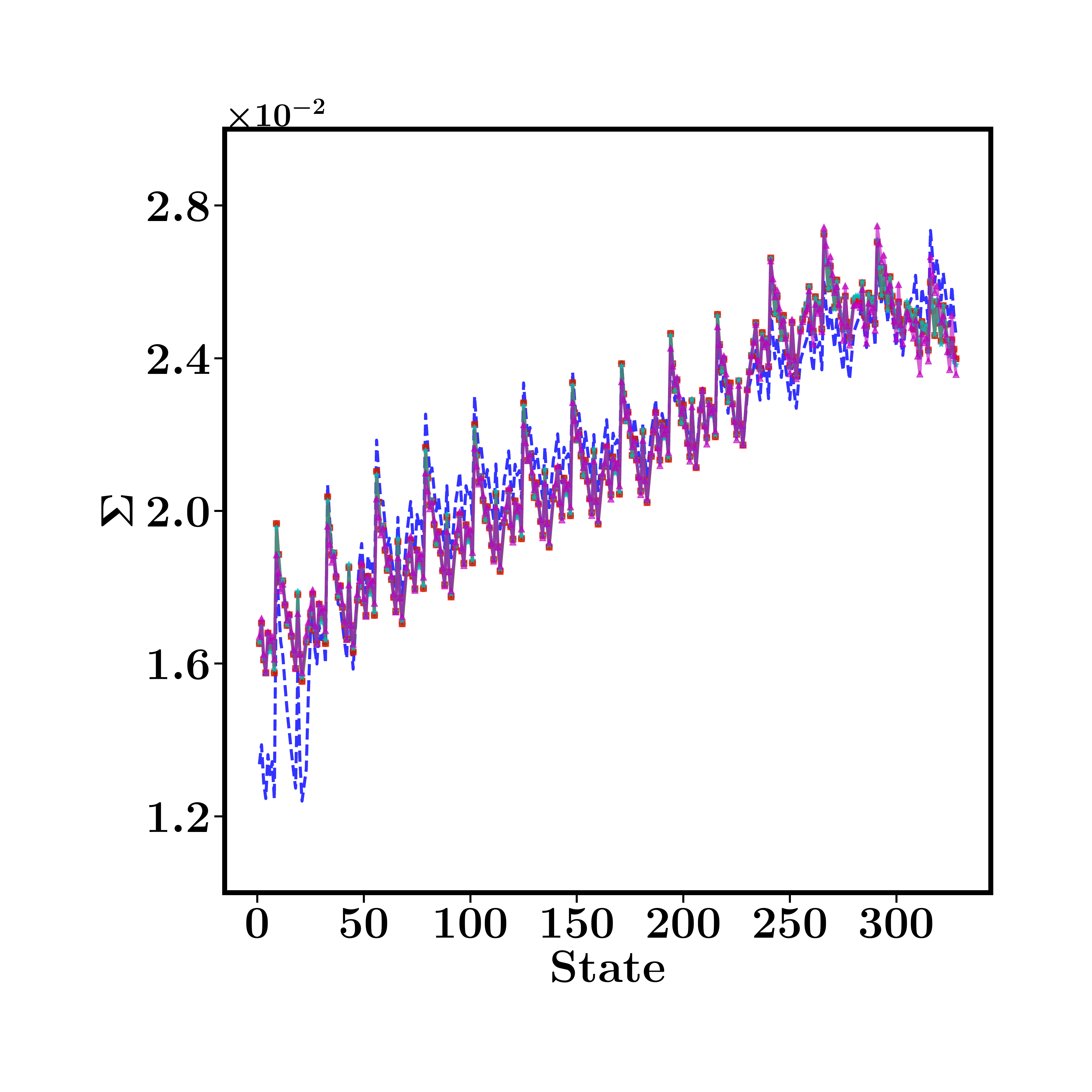}
  \subcaption{\bf Individual and base case expansion approach.}
  \label{fig:BestFit_r1_updated_1_XSF1_taylor_indiv_None}
  \end{subfigure}
  \hfill
  \begin{subfigure}[t]{0.49\textwidth}
  \centering
  \includegraphics[width=\textwidth]{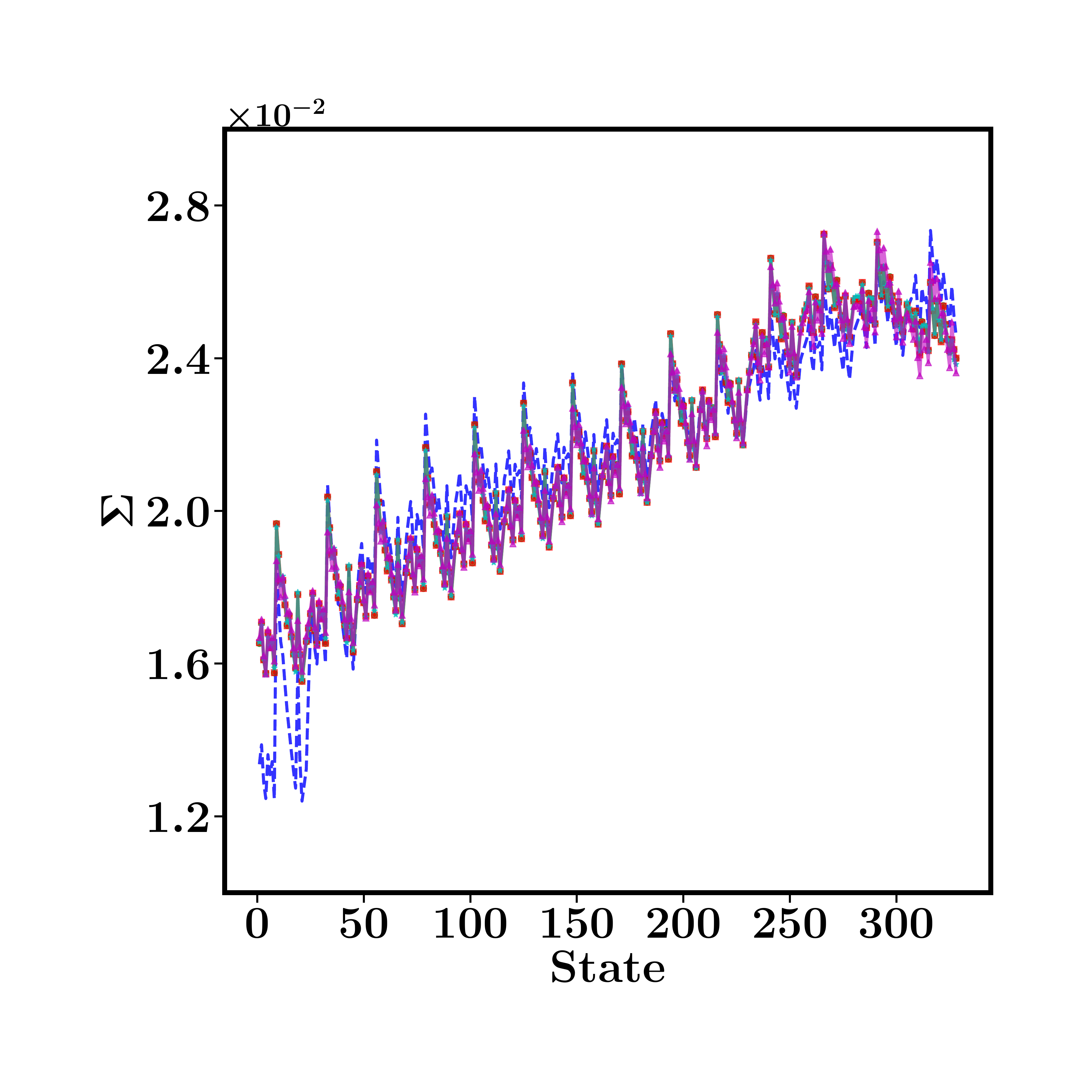}
  \subcaption{\bf Composite and base case expansion approach.}
  \label{fig:BestFit_r1_updated_1_XSF1_taylor_multi_None}
  \end{subfigure}
  \vfill
  \begin{subfigure}[t]{0.49\textwidth}
  \centering
  \includegraphics[width=\textwidth]{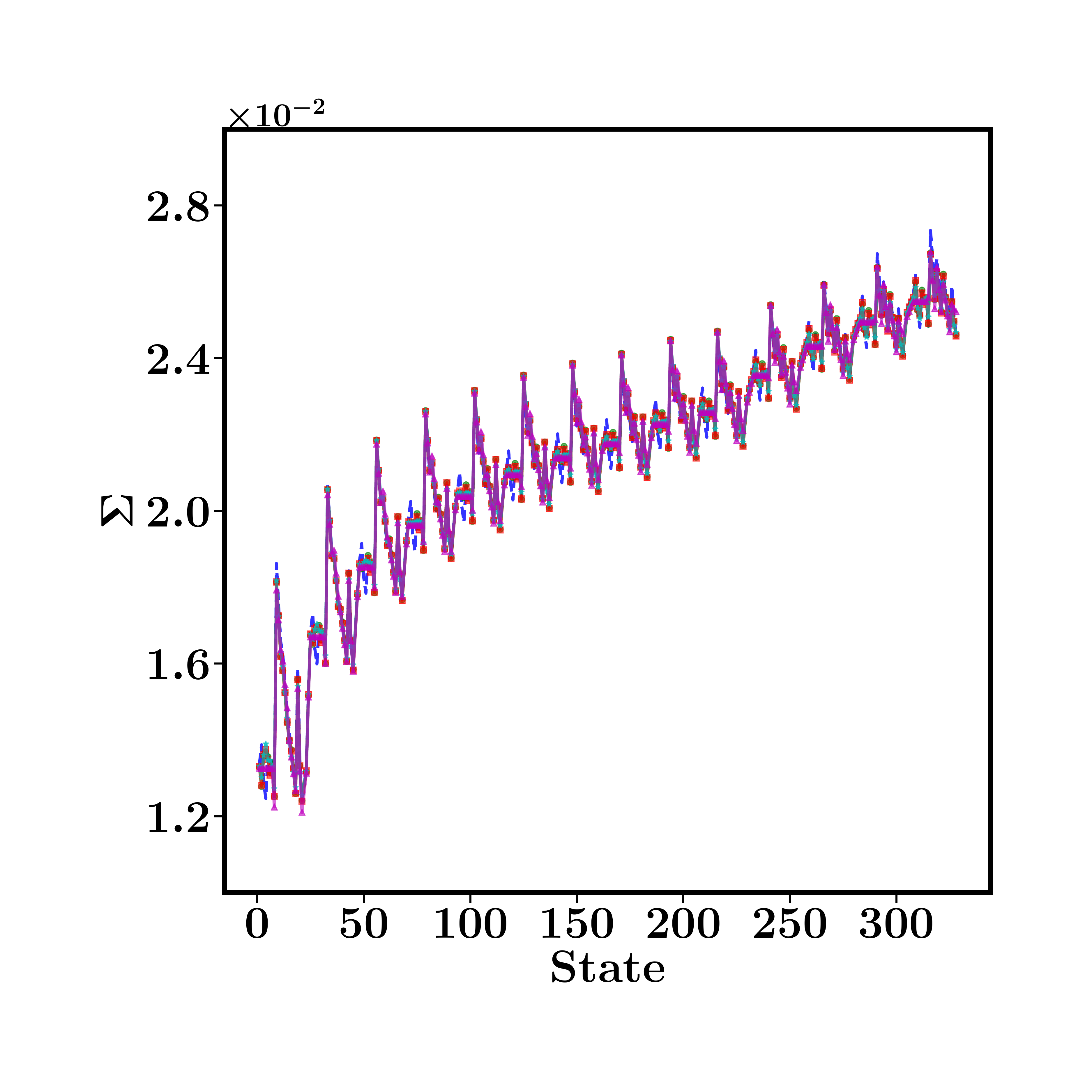}
  \subcaption{\bf Individual and burnup expansion approach.}
  \label{fig:BestFit_r1_updated_1_XSF1_taylor_indiv_BURNUP}
  \end{subfigure}
  \hfill
  \begin{subfigure}[t]{0.49\textwidth}
  \centering
  \includegraphics[width=\textwidth]{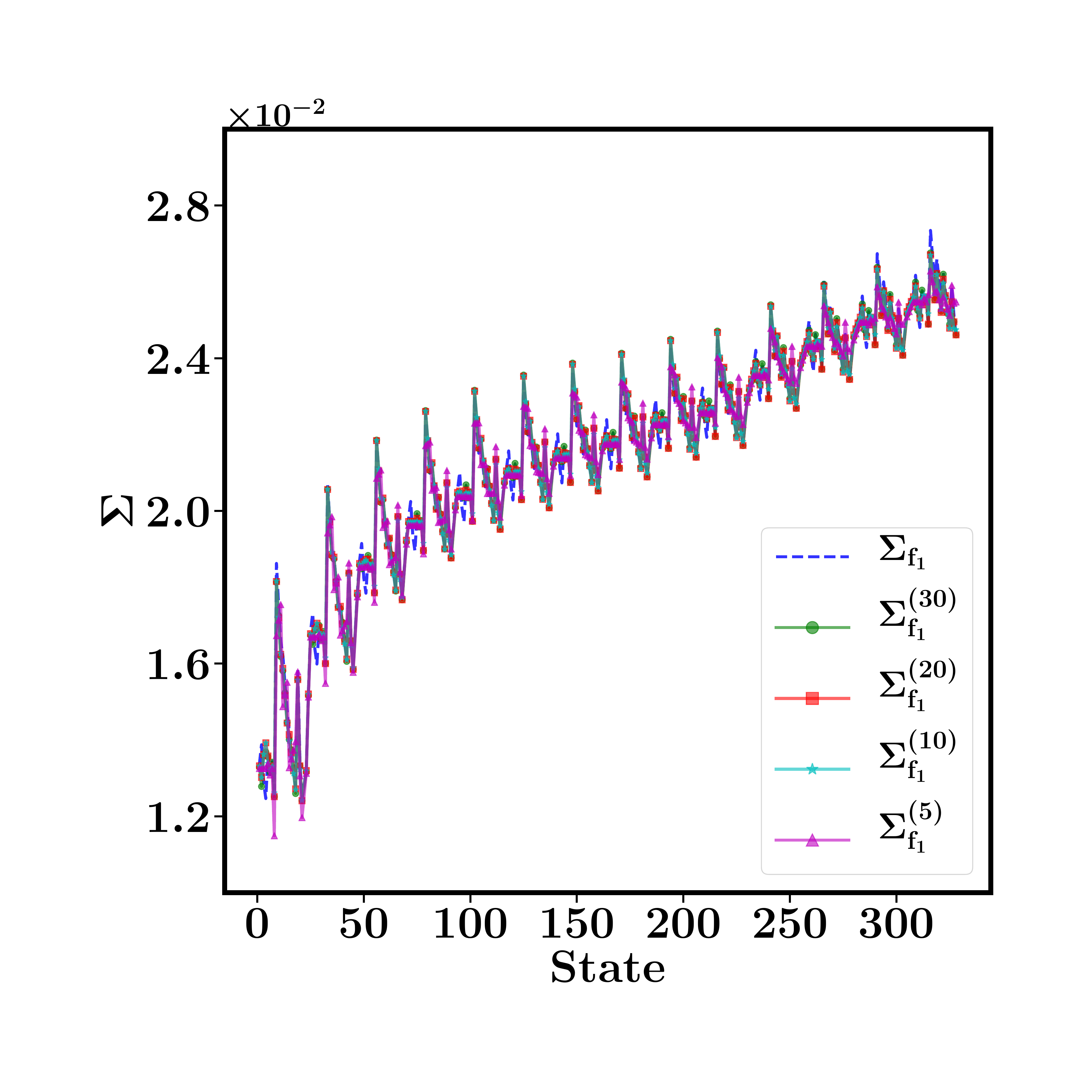}
  \subcaption{\bf Composite and burnup expansion approach.}
  \label{fig:BestFit_r1_updated_1_XSF1_taylor_multi_BURNUP}
  \end{subfigure}
  \caption{\bf Best fits of cross section over case matrix states, for decreasing number of operators for $\SIG[{\fs[1]}]$ using the individual and composite Taylor series approach, expanded around the base case and burnup states. The cross section superscript denotes the number of operators present in the model for that fit, and the dashed curve denotes the cross section data.}
  \label{fig:BestFit_r1_updated_1_XSF1_taylor_indiv_multi_BURNUP_None}
\end{figure}

In addition to the loss curves, the actual fits of the cross sections over the case matrix states, with fits for the models with $p=30,20,10,5$ operators, are shown in the fitted curves for the cross section  in \cref{fig:BestFit_r1_updated_1_XSF1_taylor_indiv_multi_BURNUP_None}. It is even more apparent from these fits that the expansion around the burnup states approach fits the oscillations of the cross sections over the case matrix much better than the base case expansion, particularly at the initial and final states.

From the step-wise regression, magnitudes of the coefficients as the number of operators decreases from $p=10,\dots,1$, are shown in the order that they are rejected from the model in Fig. \ref{fig:Coef_r1_updated_1_XSF1_taylor_indiv_multi_BURNUP_None}. Here, both the individual and composite approaches, as well as the coefficients from the expansions around the base case and burnup states, are shown separately. From these plots, as the model becomes more parsimonious the importance of each operator can be determined and how the importance fluctuates over the step-wise regression offers insight into how the model evolves with less operators. Some parameters evolve in magnitude similarly, also indicating how similar some operators are in their relevance in the model. Ultimately, the most important operators, with the loss curves, allow a resulting model to be proposed for each approach.

The summarized models for the expansions about the burnup states, showing the leading most important terms from the step-wise regression, are as follows for the individual approach for $\SIG[{\fs[1]}]$,
\begin{align}
  \SIG[{\fs[1]}] = \gamma^{\BRN} \SIG[{\fs[1]}](\BRN) ~&+~
  \gamma^{\{\ROD\}}\unidifference{\SIG[{\fs[1]}](\BRN)}{\ROD} \Delta \ROD ~+~ O(\Delta \var)  \\ &+~
  \frac{1}{2} \gamma^{\{\RHOM \TF\}} \difference{\SIG[{\fs[1]}](\BRN)}{\RHOM, \TF } \Delta \RHOM \Delta \TF ~+~  O(\Delta \var\Delta\var[\varindextwo]), \nonumber
\end{align}
and for the composite approach for arbitrary cross section and group,
\begin{align}
  \SIG[\funcindex_{g}] = \gamma^{\BRN} \SIG[\funcindex_{g}](\BRN) ~&+~
  \gamma^{\{\RHOM\}} \unidifference{\SIG[\funcindex_{g}](\BRN)}{\RHOM} \Delta \RHOM + O(\Delta \var) \\ &+~
  \frac{1}{2}\gamma^{\{\TM \TF\}} \difference{\SIG[\funcindex_{g}](\BRN)}{\TM,\TF} \Delta \TM \Delta \TF ~+~ O(\Delta \var\Delta\var[\varindextwo]). \nonumber
\end{align}
Each of the quantities of interest are shown to require few operators before regression loss significantly increases. The loss curves as the model becomes more parsimonious, are also lower for the burnup state expansion than the base case expansions, indicating the influence of the burnup as an important time-like quantity. 

The individual step-wise regression approach is also shown to have much smoother loss curves than the composite approach, demonstrating effects of overfitting, and that the cross section independent model must be augmented with knowledge of other parameters, assuming this independence from the type of cross section is a valid assumption.

\begin{figure}[ht]
  \centering
  \begin{subfigure}[t]{0.6\textwidth}
  \centering
  \includegraphics[width=\textwidth]{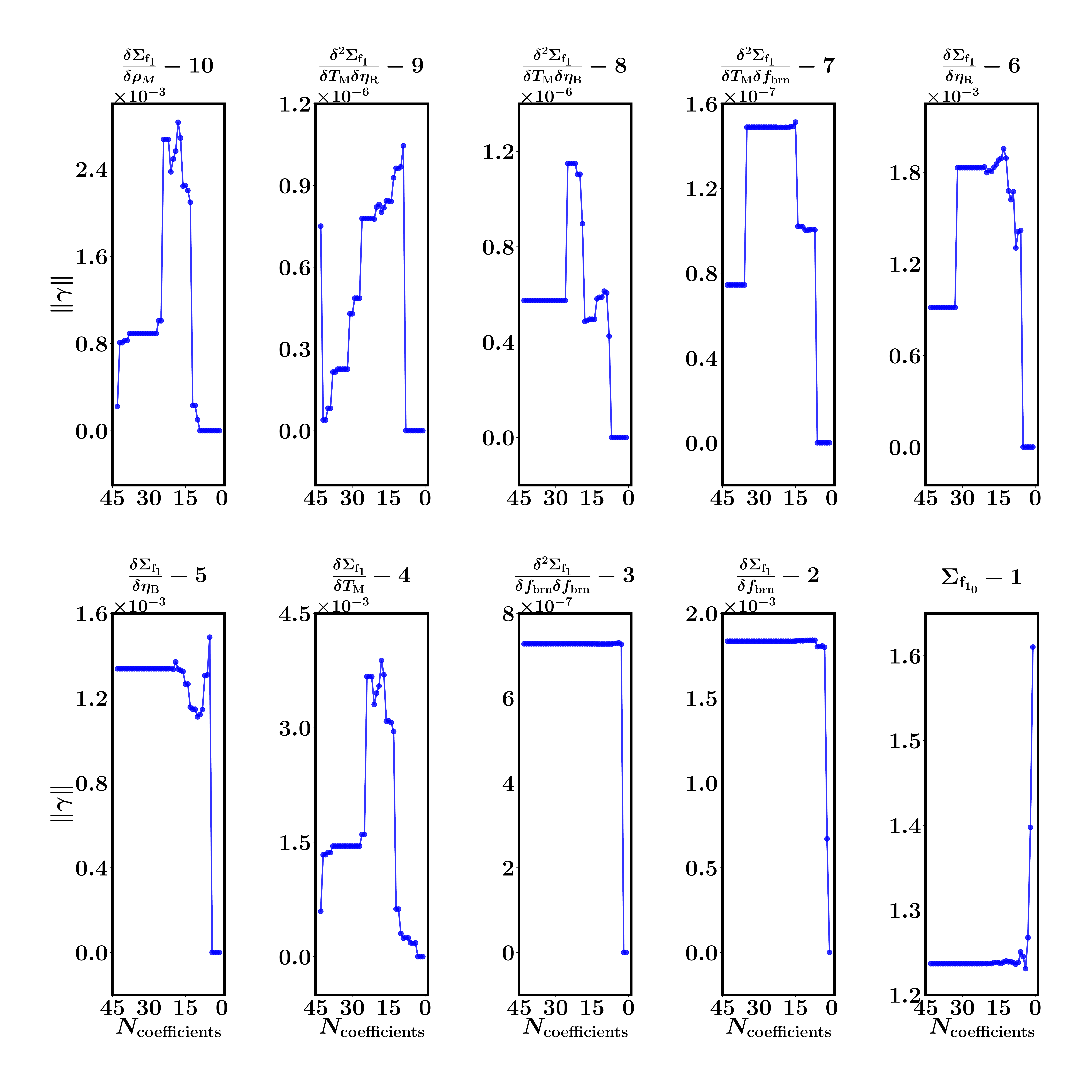}
  \subcaption{\bf Individual and base case expansion approach.}
  \label{fig:Coef_r1_updated_1_XSF1_taylor_indiv_None}
  \vspace*{0.5cm}
  \end{subfigure}
  \begin{subfigure}[t]{0.6\textwidth}
  \centering
  \includegraphics[width=\textwidth]{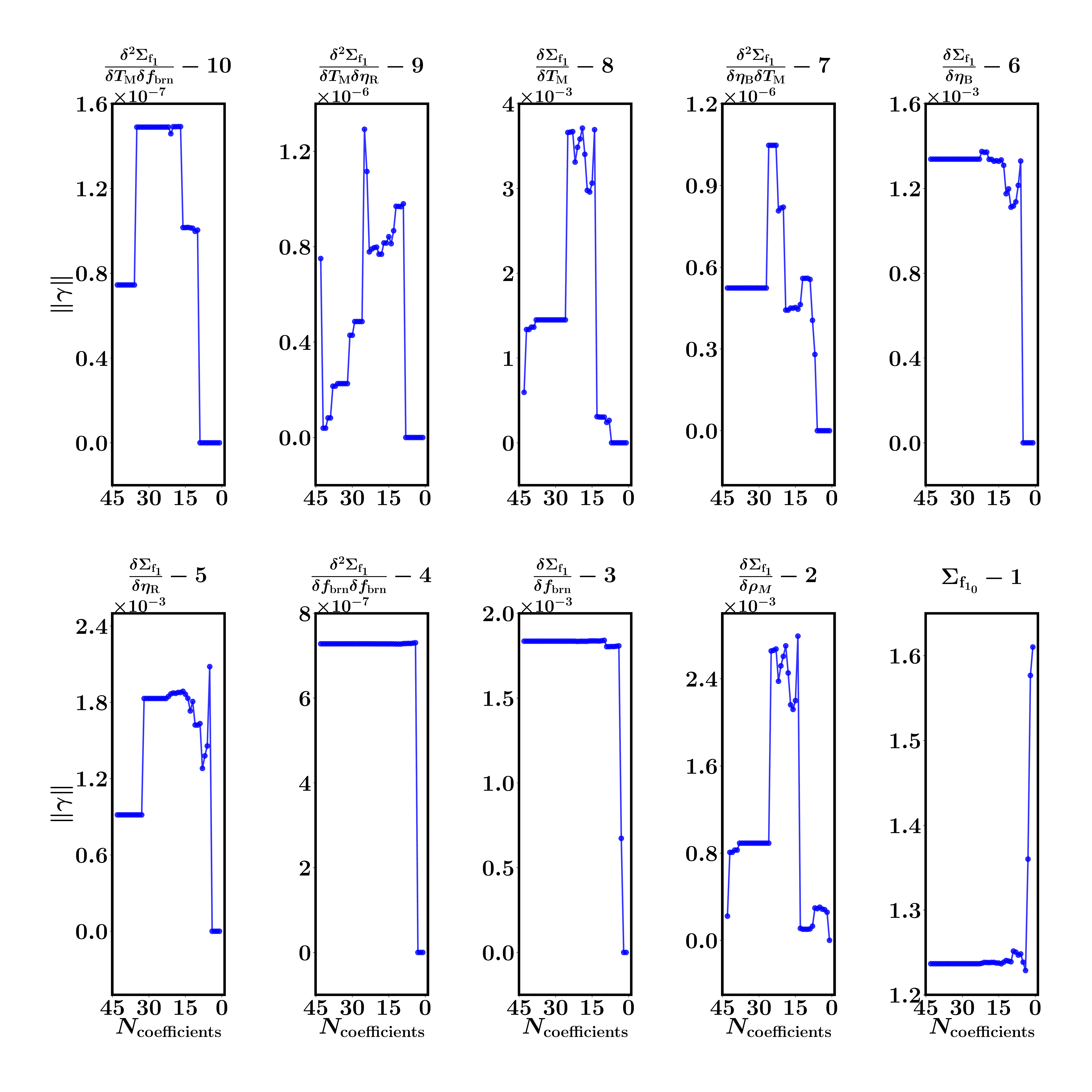}
  \subcaption{\bf Composite and base case expansion approach.}
  \label{fig:Coef_r1_updated_1_XSF1_taylor_multi_None}
  \end{subfigure}
  \end{figure}
  \begin{figure}[ht]\ContinuedFloat
  \centering
  \begin{subfigure}[t]{0.6\textwidth}
  \centering
  \includegraphics[width=\textwidth]{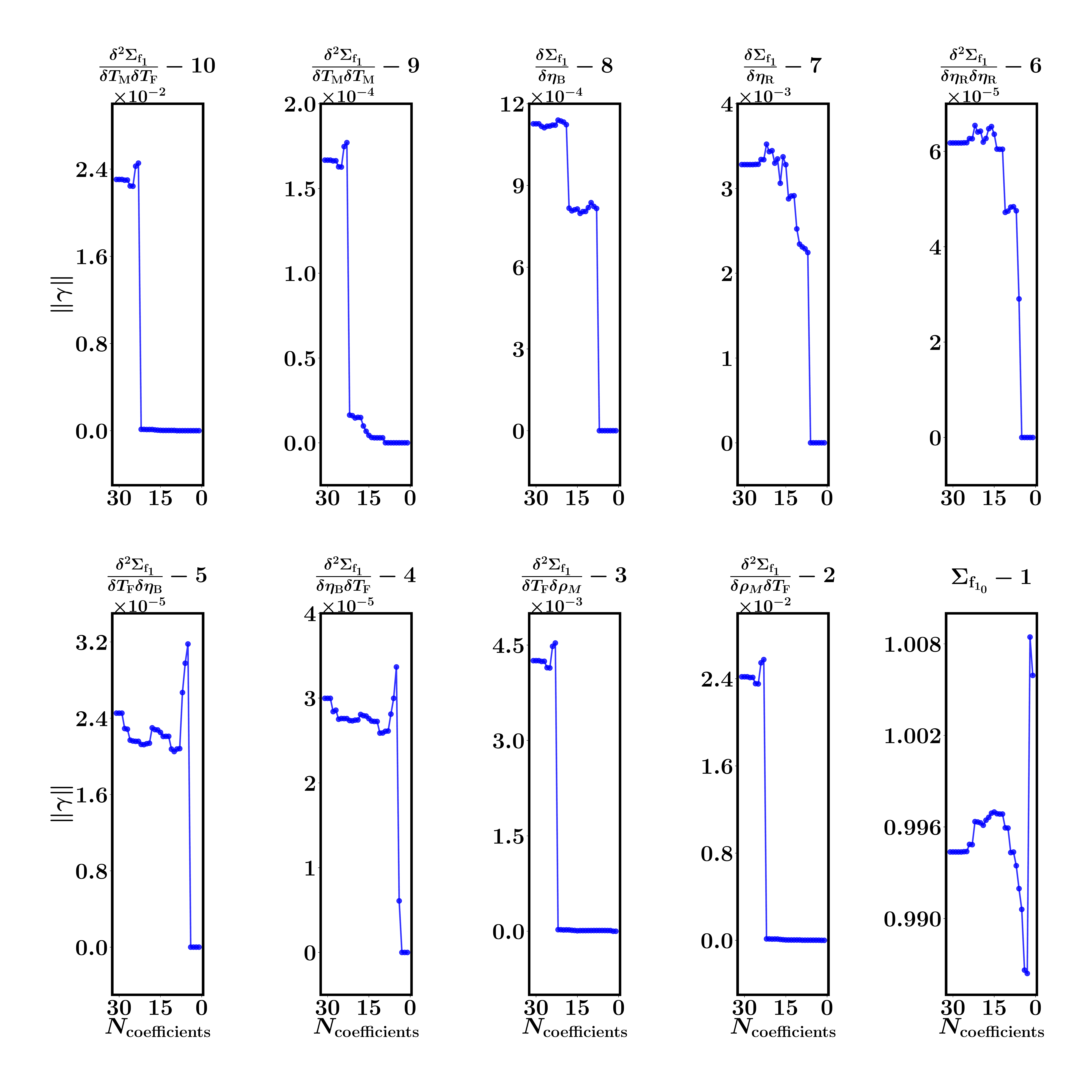}
  \subcaption{\bf Individual and burnup expansion approach.}
  \label{fig:Coef_r1_updated_1_XSF1_taylor_indiv_BURNUP}
  \vspace*{0.5cm}
  \end{subfigure}
  \begin{subfigure}[t]{0.6\textwidth}
  \centering
  \includegraphics[width=\textwidth]{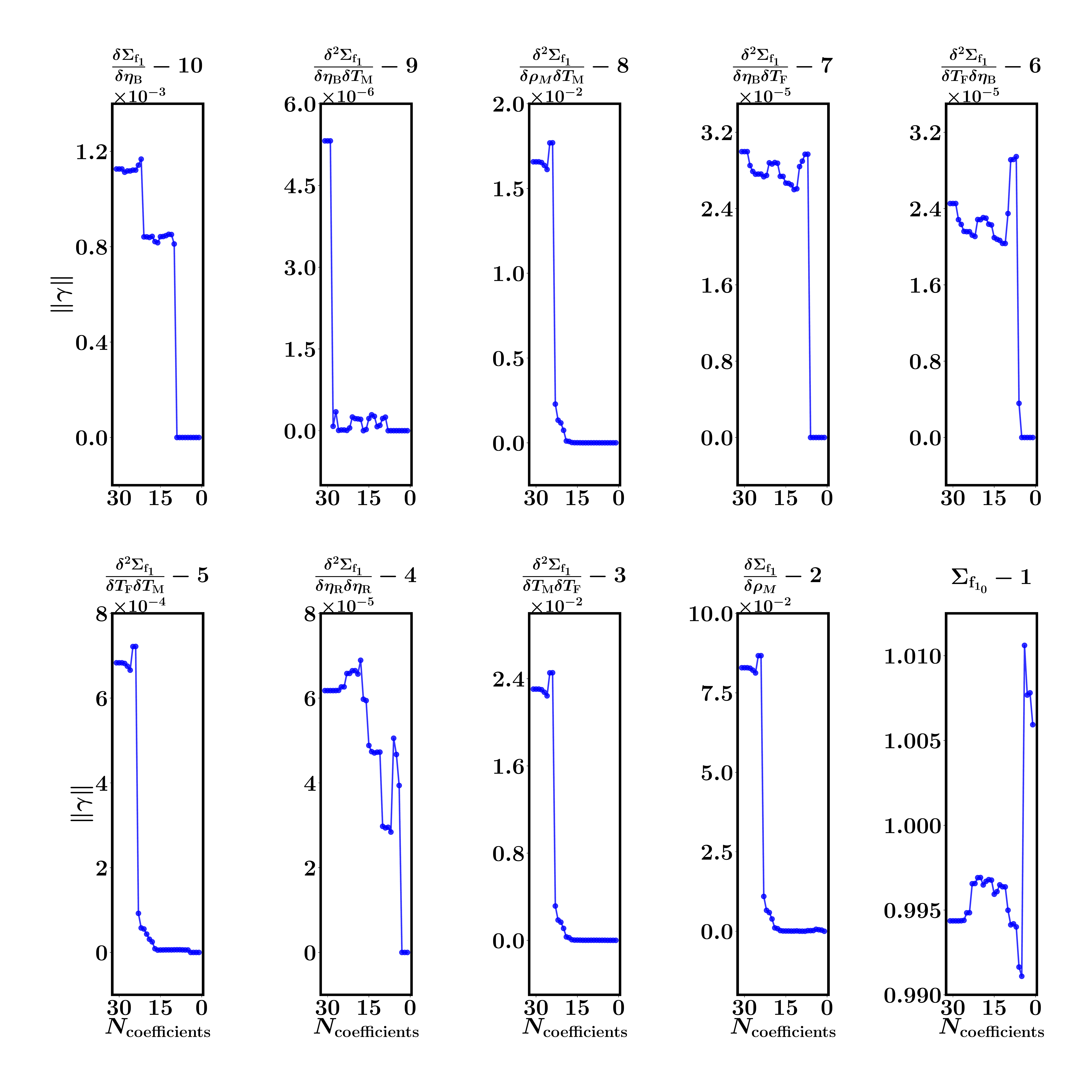}
  \subcaption{\bf Composite and burnup expansion approach.}
  \label{fig:Coef_r1_updated_1_XSF1_taylor_multi_BURNUP}
  \end{subfigure}
  \caption{\bf Coefficient magnitudes for decreasing number of operators for $\SIG[{\fs[1]}]$ using the individual and composite Taylor series approach, expanded around the base case and burnup states. The operator and integral value in each subfigure title denotes the number of operators present in the model when the given operator is rejected.}
  \label{fig:Coef_r1_updated_1_XSF1_taylor_indiv_multi_BURNUP_None}
\end{figure}


\section{Conclusions} 
Models for the parameterization of cross sections at different conditions along a history of burnup values are shown to be adaptable to the GTA. Initial numerical results corroborate the intuition of expert reactor physicists that have developed the models currently used. The choice of a Taylor series expansion, with important operators selected via step-wise regression, is an effective basis to model cross sections from the low dimensional set of state parameters. Individual or general models for cross sections can be found, and given the burnup history, the burnup as a base time-like quantity yields low fitting losses with a significantly truncated expansion. Thus, the GTA should lend a more general, yet rigorous, methodology to develop parameterized cross section models, and future applications include computing full reduced order models, and predicting cross section data for reactor conditions not usually considered in conventional case matrix studies.
Future communications will include details omitted here for brevity.

\section*{Acknowledgements} 
We acknowledge the support of NSF DMREF grant \#1729166 (MD and KG). This work was also partially supported by an interdisciplinary mini-grant from the Fastest Path to Zero Initiative. \\(\url{https://fastestpathtozero.umich.edu/}).

\setlength{\baselineskip}{12pt}
\bibliographystyle{main}
\bibliography{main}

\end{document}